\documentclass[12pt]{article}
\usepackage{pic03}
\usepackage{hyperref}
\usepackage{url}
\usepackage{graphicx}

\begin{document}

\title{\bf RECENT RESULTS ON LIGHT MESON PHYSICS}
\author{
Cesare Bini    \\
{\em Dipartimento di Fisica, Universit\`a ``La Sapienza'' e INFN, Roma, Italy}}
\maketitle

%
%
%
%
%
%
\vspace{2.5cm}
%

\baselineskip=14.5pt
\begin{abstract}
Some recent results on light meson physics are reviewed. The new evidence of 
low mass scalar mesons together with the improved measurement of the $\phi$ radiative
decays in scalar mesons, give new insight into the nature and the structure of the
scalar spectrum. The evidence of new states with a mass close to twice the
proton mass, and a new
analysis of the gluonium content of $\eta'$ are also discussed.   

\end{abstract}
\newpage

\baselineskip=17pt

\section{Introduction}
The physics of light mesons has received in the last years several inputs
from fixed target experiments with hadrons or photons beams 
and from $e^+e^-$ experiments. The main aim of the published 
analyses is to clarify and to understand the very
rich low mass (say below 2 GeV) mesons spectrum. In particular it is
considered of primary
importance to find signatures of physical states that cannot be
interpreted as bound states of a quark and an antiquark
($q\overline{q}$). In fact QCD and QCD inspired models predict the
existence of
bound states of gluons (glueballs), of quark composites with 
more than 2 quarks (4-quark states
$q\overline{q}q\overline{q}$) and of bound states of 
quarks and gluons (hybrid states
$q\overline{q}$+gluon). All these states are generally called exotics.

In this paper some recent results in this field are presented
and discussed.

\section{Results on Scalar Mesons}
The scalar meson spectrum is  
a place where the so-called exotics are particularly
searched. In fact the number of experimentally found 
physical states
with scalar quantum numbers (namely $J^{PC}=0^{++}$) is larger than
expected. So, while it is easy
to accommodate pseudo-scalar ($J^{PC}=0^{-+}$) and vector
($J^{PC}=1^{--}$) mesons in $q\overline{q}$  SU(3) nonets,
this is not true for scalar mesons. The list of the presently known 
scalar states is given in
Tab.\ref{scalars}. The states of
the lowest mass $q\overline{q}$ nonet according to the Particle Data
Group (PDG in the following \cite{PDG}) are indicated in the table. 
As we shall see in the following the scalar nonet  
identification is still controversial and the PDG choice is only one among
the possible interpretations.

New recent insight on light scalar mesons come from high statistics
studies of
three-body D-mesons and $J/\psi$ decays, and from $\phi$ radiative decays.
\subsection{New evidence of low mass scalar states: $\sigma$ and $\kappa$}
Three-body D-mesons and $J/\psi$ decays are analysed looking at
two-dimensional Dalitz plot distributions\cite{E791,Focus,Bes1,Cleo}. 
This analysis has been applied to the following Dalitz plots:
\begin{itemize}
\item{$D^+(D^+_{(s)})\rightarrow\pi^+\pi^+\pi^-$ decays where any $\pi^+\pi^-$ pair is
sensitive to scalar isoscalar intermediate states;}
\item{$D^+\rightarrow K^+\pi^+\pi^-$, $D^0\rightarrow K^-\pi^+\pi^0$ and 
$D^0\rightarrow K_S\pi^+\pi^-$ decays where any $K\pi$ pair is sensitive to
isospin 1/2 scalar intermediate states;}
\item{$J/\psi\rightarrow K^{*0}K^+\pi^-$ also sensitive to $K\pi$ isospin 1/2 states; }
\end{itemize}

\begin{figure}[htb]
\includegraphics[width=13cm]{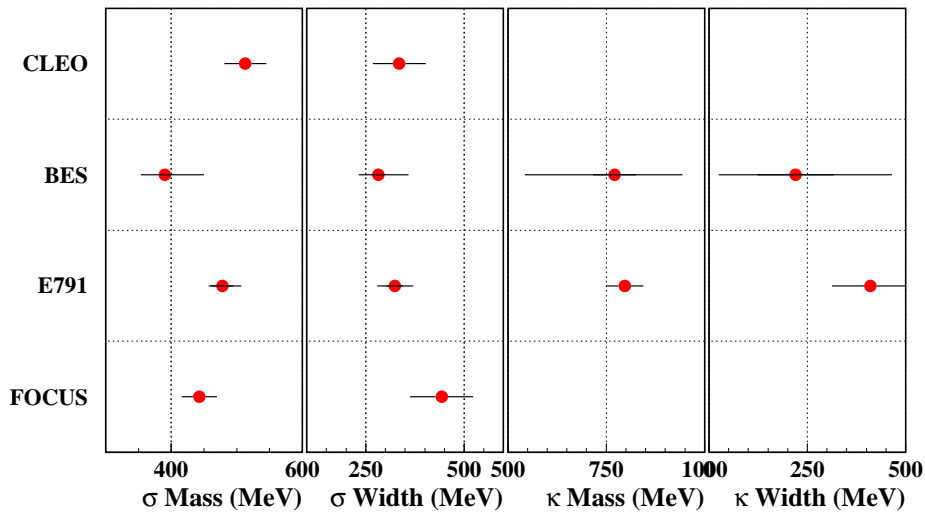}
 \caption{\it
      Summary of recent results on masses and widths of $\sigma$ and $\kappa$.
    \label{sigmakappa} }
\end{figure}

\begin{table}
\centering
\caption{ \it List of scalar states with mass below 1.8 GeV ordered in increasing
  mass. The states in parentheses are included in the $q\overline{q}$
  lowest mass nonet by PDG.}
\vskip 0.1 in
\begin{tabular}{|c|c|c|} \hline
 I=0 & I=1/2 & I=1 \\
\hline
 $f_0(400-1200)$ ($\sigma$) & $\kappa(700)$ & $a_0(980)$ \\
 $f_0(980)$ & $[K^*_0(1430]$ & $[a_0(1450)]$ \\
 $[f_0(1370)]$ &  &  \\
 $f_0(1500)$ &  &  \\
 $[f_0(1710)]$ &  &  \\
\hline
\end{tabular}
\label{scalars}
\end{table}
In order to fit the Dalitz plots two low
mass broad states are introduced: a isoscalar $\sigma$ and an isospin 1/2
$\kappa$. Each experiment suggests values for masses and widths of these
states. These values are summarised in Fig.\ref{sigmakappa}. The values 
are in good agreement and the general
indication is for an isoscalar state with a mass of 470 MeV and a
width of 340 MeV and a isospin 1/2 state at 800 MeV with a width of 410 MeV.  
The broad $\sigma$ state has been also introduced several times in the
past, and its existence has always been considered controversial.
Notice that the approach used by these experiments
to extract the informations on the scalar states, has been
recently criticised \cite{Focus,AS2}.

\subsection{$f_0(980)$ and $a_0(980)$ in $\phi$ radiative
decays} 
The $\phi$(1020) is interpreted as an almost pure $s\overline{s}$
state. Due to the OZI rule the decays in final states not containing the
$s$-quark are suppressed. In this context
the branching ratio of a $\phi$ going to a meson plus a  
photon depends on the $s$-quark content of the meson itself.

\begin{figure}[htbp]
  \centerline{\hbox{ \hspace{0.2cm}
    \includegraphics[width=6.0cm]{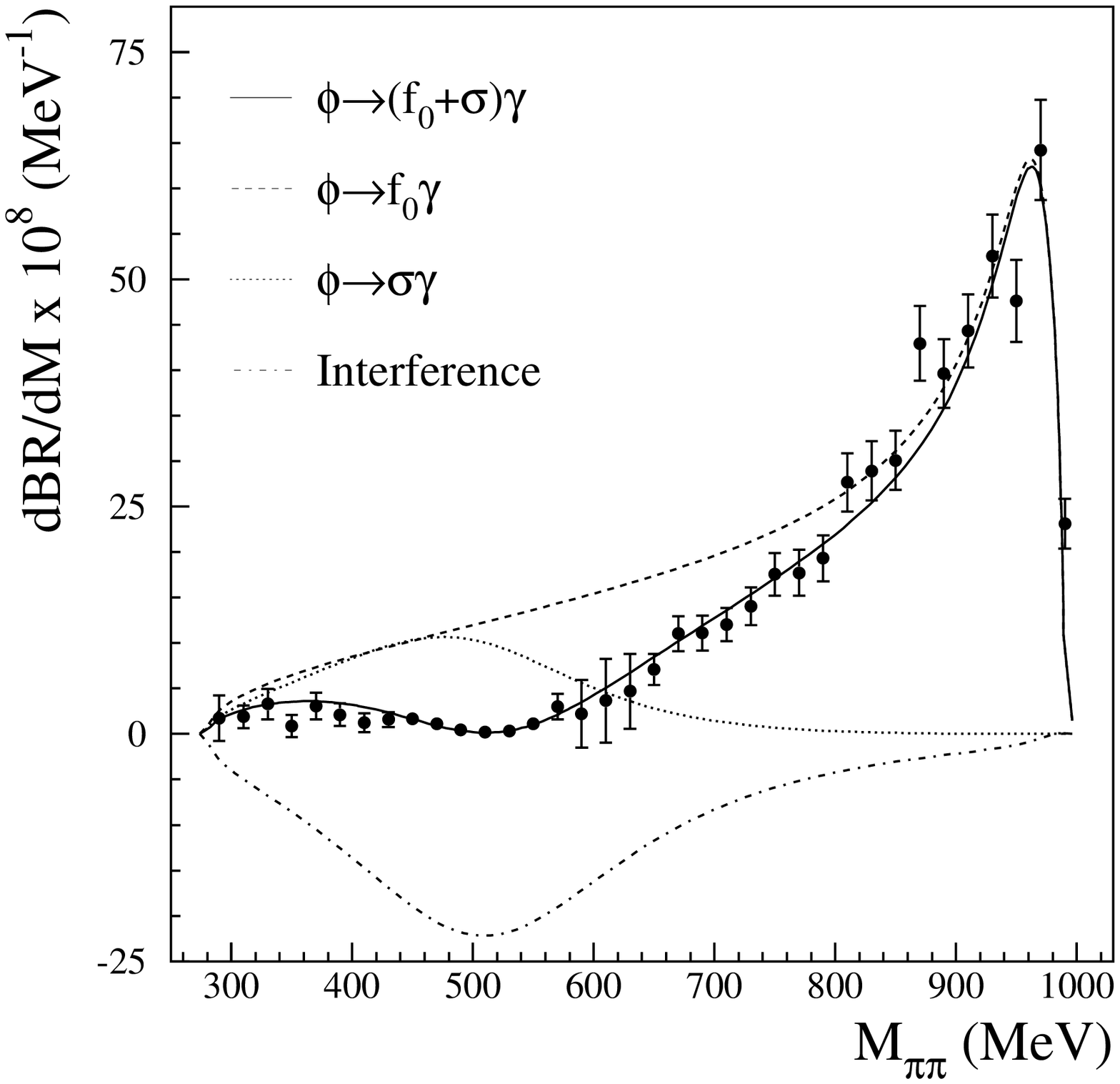}
    \hspace{0.3cm}
    \includegraphics[width=7.0cm]{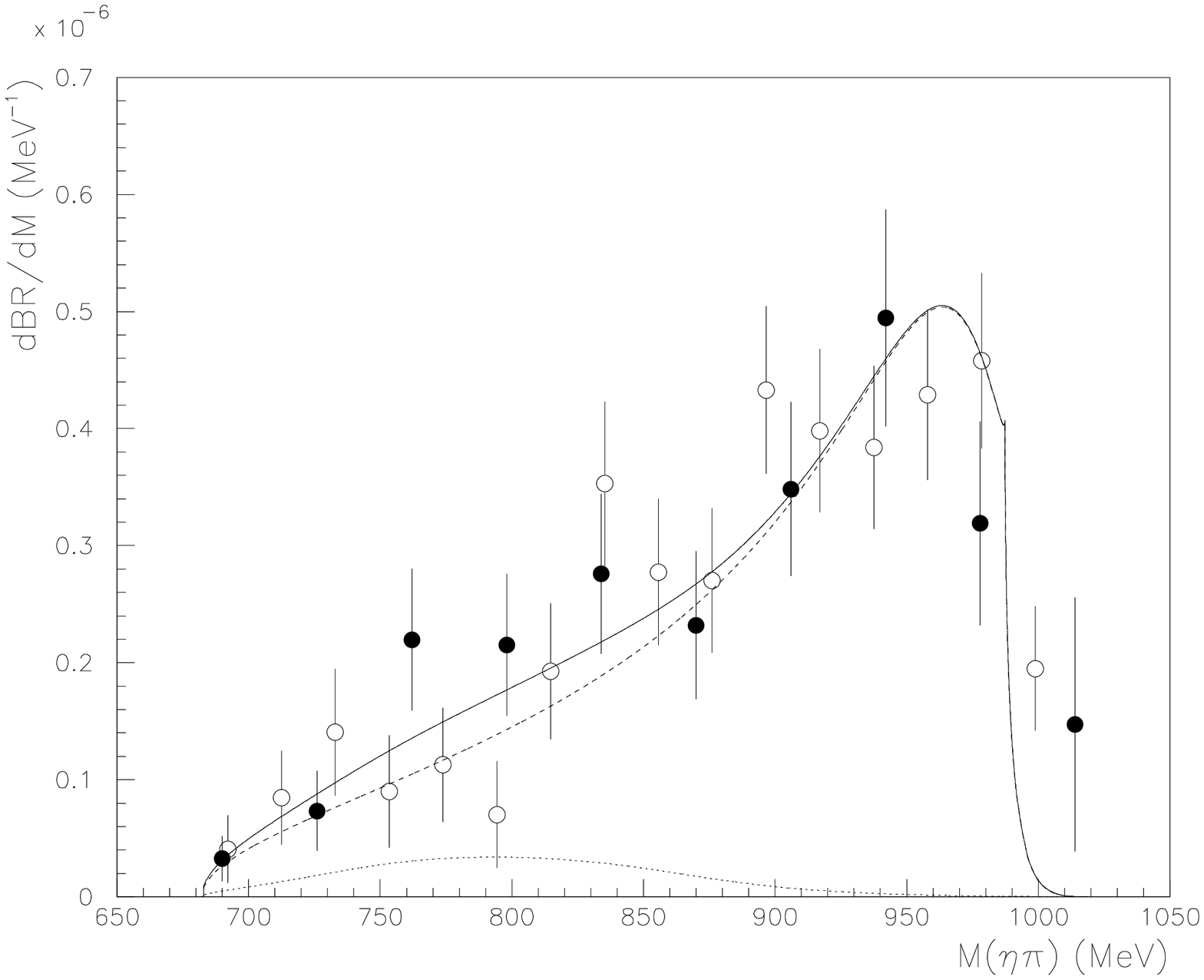}
    }
  }
 \caption{\it
    Mass spectra of $\pi^0\pi^0$ (left) and $\eta\pi^0$ (right) final states
    from KLOE with fit superimposed.  
    \label{KLOE} }
\end{figure}

At a $\phi$ factory ($e^+e^-$ collider at 1020 MeV centre of mass
energy) the scalar states $f_0$(980), $a_0^0$(980) and $\sigma$ are
accessible through the following final states:
\begin{itemize}
\item{ $\phi\rightarrow\pi^+\pi^-\gamma$ and
$\phi\rightarrow\pi^0\pi^0\gamma$ where the two pions are in a scalar
isoscalar state (IJ$^{PC}$=00$^{++}$);}
\item{ $\phi\rightarrow\eta\pi^0\gamma$ where the $\eta\pi^0$ system is a
scalar isovector state (IJ$^{PC}$=10$^{++}$);}
\end{itemize} 
KLOE at the $\phi$-factory DAFNE at Frascati has analysed the
decays $\phi\rightarrow\pi^0\pi^0\gamma$ \cite{KLOE1} 
and $\phi\rightarrow\eta\pi^0\gamma$ \cite{KLOE2}
\footnote{The first results on these decays have been obtained few years
  ago by
SND and CMD-2 at VEPP-2M \cite{SND,CMD-2},
Novosibirsk. KLOE has improved by more than a factor 10 in statistics
these results.} 
The observed branching ratios are both of the order $10^{-4}$: 
$B.R.(\phi\rightarrow\pi^0\pi^0\gamma)=(1.08\pm0.05_{stat}\pm0.03_{syst})\times
10^{-4}$, 
$B.R.(\phi\rightarrow\eta\pi^0\gamma)_1=(0.85\pm0.05_{stat}\pm0.06_{syst})\times
10^{-4}$ and $B.R.(\phi\rightarrow\eta\pi^0\gamma)_2=(0.80\pm0.06_{stat}\pm0.04_{syst})\times
10^{-4}$,
where the $B.R._{1,2}$ for the $\eta\pi^0\gamma$ decay, refer to 2
different $\eta$ final states (namely $\eta\rightarrow\gamma\gamma$ and
$\eta\rightarrow\pi^+\pi^-\pi^0$). 
The mass spectra of the $\pi^0\pi^0$ and $\eta\pi^0$ systems are 
shown in Fig.\ref{KLOE}. The spectra
are fitted with a parametrisation based on the 
kaon-loop model \cite{AcIv} and turn out to
be dominated by the scalar particle production, any other contribution
being  negligible\footnote{Since the masses of $f_0$ and $a_0^0$ are very
close to the mass of the $\phi$ the typical Breit-Wigner shape
is distorted. It is essentially multiplied by a $p_{\gamma}^3$ 
factor, where $p_{\gamma}$ is the momentum of the radiated photon.} 
In the case of the $\pi^0\pi^0$ spectrum a good fit is obtained only including a 
$\sigma$ with the same parameter and mass shape found by E791 \cite{E791}, 
negatively interfering with the $f_0$ signal.
If the quoted branching ratios are completely attributed to $\phi\rightarrow$ 
Scalar+$\gamma$ decays, as it appears from the analysis of the spectra, 
the standard $q\overline{q}$ interpretation of $f_0$ and $a_0$ is 
in trouble. In fact the only possible quark compositions of $f_0$ 
and $a_0$ compatible with the observed mass degeneracy is:
$f_0\sim(u\overline{u}+d\overline{d})$
$a_0^0\sim(u\overline{u}-d\overline{d})$
requiring branching ratios of the $\phi$ to $f_0\gamma$ or
$a_0\gamma$ of the order of $10^{-6}$ (due to OZI suppression), 
that is 2 order of magnitude lower than the observed 10$^{-4}$. 

On the other hand, this large branching ratio could be well explained in
the context of 
$q\overline{q}q\overline{q}$ model where the $f_0$ and $a_0$ quark
structure should be:
$f_0\sim(u\overline{u}+d\overline{d})s\overline{s}$
$a_0^0\sim(u\overline{u}-d\overline{d})s\overline{s}$
both with an explicit $s\overline{s}$ pair. In this case the branching
ratios should be of the order of $10^{-4}$.

KLOE data on $f_0$ have been also 
analysed with an approach based on the K-matrix
\cite{BogPen} aiming to get a
model-independent conclusion. The result is
dependent on the details of K-matrix used, but indicates that, in contrast
to kaon-loop model result, only a part of the spectrum should be
attributes to the $f_0$ production.

KLOE has now collected about 500 pb$^{-1}$ while the results
shown here corresponds to only 16 pb$^{-1}$. Higher accuracy results are
expected soon, including also the decay $\phi\rightarrow\pi^+\pi^-\gamma$.

\subsection{Possible scenarios}
\begin{figure}[htbp]
  \centerline{\hbox{ \hspace{0.2cm}
    \includegraphics[width=4.5cm]{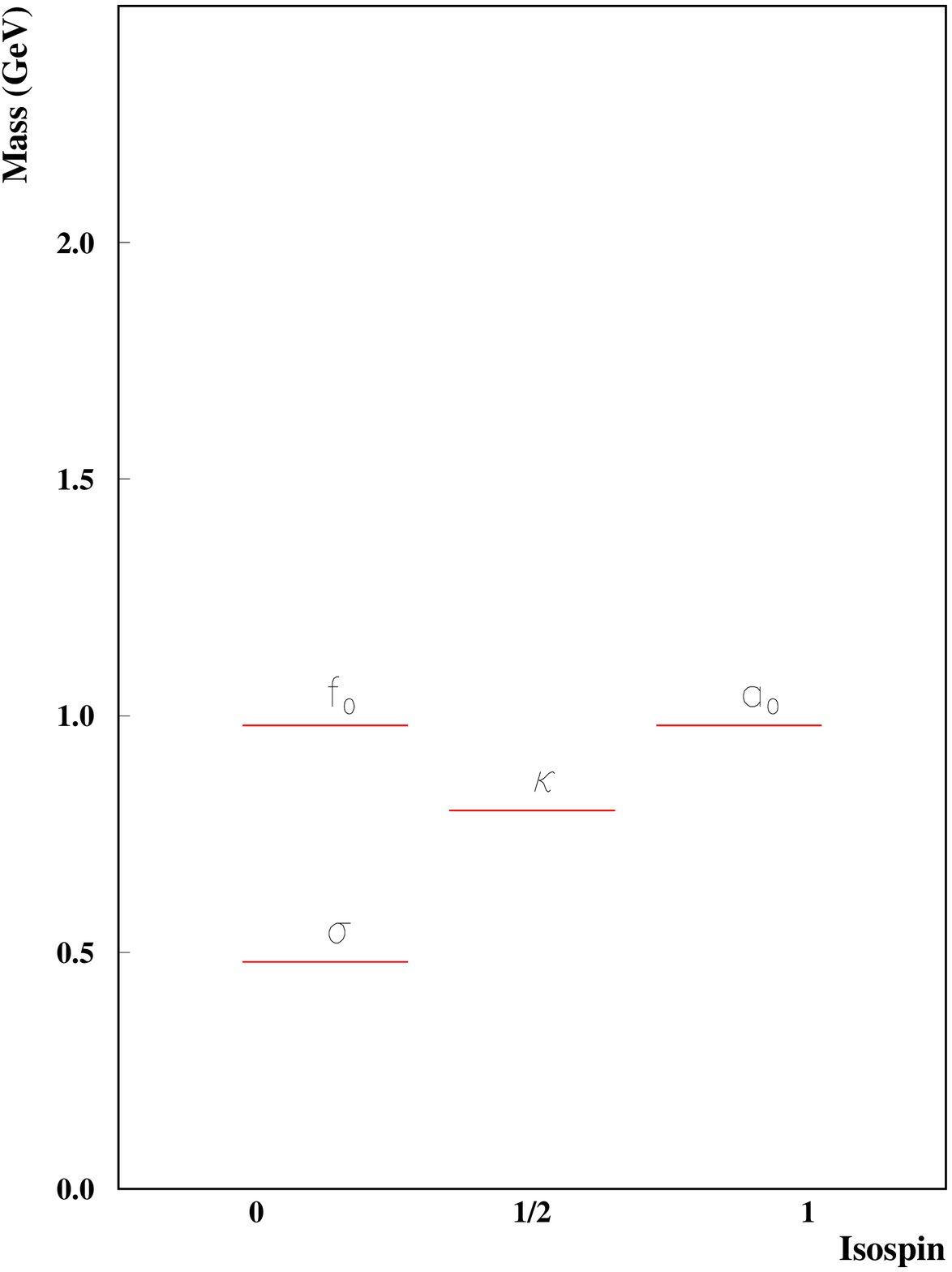}
    \hspace{0.3cm}
    \includegraphics[width=4.5cm]{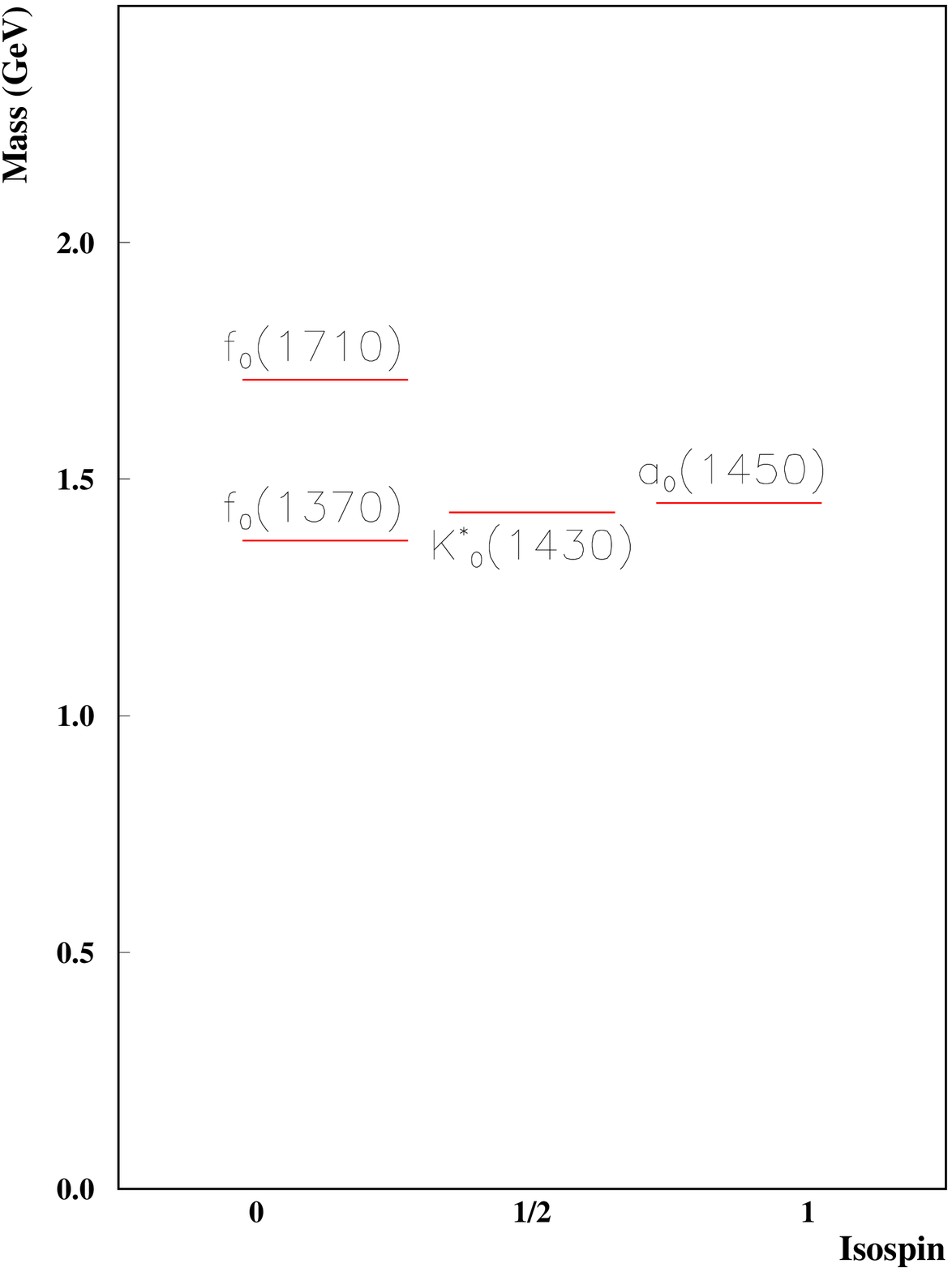}
    \hspace{0.3cm}
    \includegraphics[width=4.5cm]{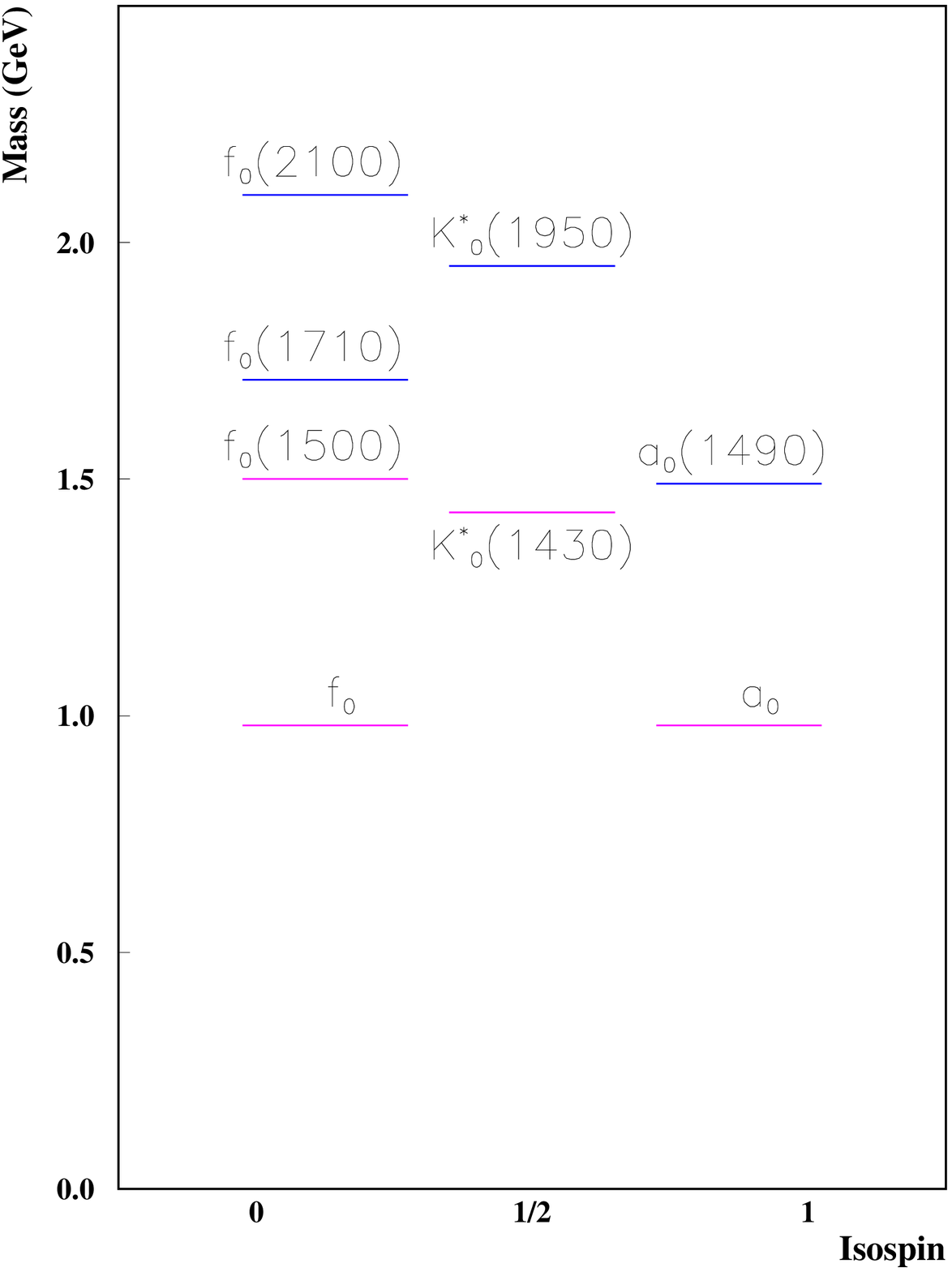}
    \hspace{0.2cm}
    }
  }
 \caption{\it
      The schemes 1 and 2 correspond to an inverted
      spectrum as suggested by the 4 quark model and to the PDG spectrum.
      The $f_0(1500)$ is excluded and could be the glueball. In scheme 3 the
      scalars are grouped in 2 $q\overline{q}$ nonets (including a couple of
      higher mass states). In this scheme the $f_0(1370)$ is excluded.
    \label{scenarios} }
\end{figure}

Looking at the scalar meson spectrum, 2 possible scenarios emerge, indicated
in Fig.\ref{scenarios}. The first scenario assumes that $\sigma$ and
$\kappa$ are real physical states. In this case there should be a lowest
mass nonet where the states are "mass inverted" as expected for a 4 quark
nonet \cite{Jaffe}, and a higher mass nonet with the PDG states. In this case only one
state, the $f_0(1500)$ should remain out, and it could be of gluonium
origin. The second scenario assumes that $\sigma$ and $\kappa$ are not
real states and gives two standard nonets \cite{Klempt}
(the second one could be
the first radial excitation of the ground nonet). This scheme requires no
$s$-quark content for $f_0(980)$ and $a_0(980)$ and so, it contradicts the
current interpretation of $\phi$ radiative decays results.

\section{New states close to $2M_{N}$.}
A second interesting subject is the recent claim of new states with an
energy very close to twice the mass of the nucleon $2M_{N}$. 

\subsection{Evidence of a dip in 6 pion diffractive photo-production}
\begin{figure}[htb]
\centering
\includegraphics[width=8cm]{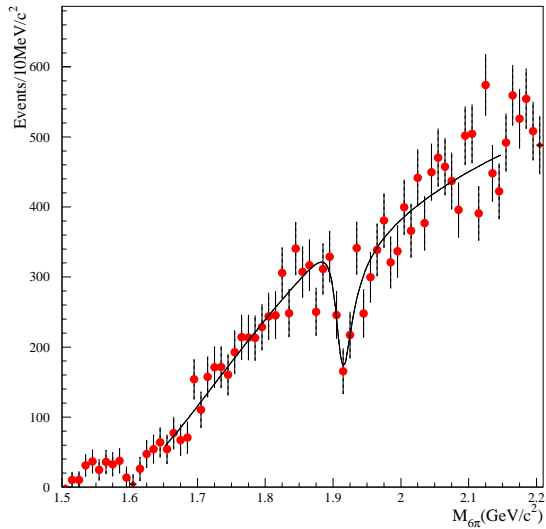}
 \caption{\it
$3\pi^+3\pi^-$ invariant mass distribution for diffractive events after
correction for acceptance and unfolding of the resolution. The curve
superimposed is the best fit. Data and fit from E687.
    \label{e687} }
\end{figure}

The E687 experiment at Fermilab has analysed the data on 
diffractive photo-production of 6
pions, $3\pi^+ 3\pi^-$ \cite{E687}. 
This analysis was motivated by 
statistically limited indications from $e^+e^-$ experiments \cite{DM2,Fenice}
of a structure in the region of $2M_N$.
In order to select final state with vector quantum numbers, as in $e^+e^-$
collisions, diffractive events are selected. This is done requiring the
square of the total transverse momentum $P^2_T$ to be below 0.040 GeV$^2$. 
The invariant mass distribution of
the 6 pion system for diffractive events shown in
Fig.\ref{e687} has a clear dip at
about 1910 MeV (that is 30 MeV above  $2M_N$) while the distribution for
non-diffractive events
doesn't show the dip. The fit shown is a coherent sum of a relativistic
Breit-Wigner resonance with free mass and width and a diffractive continuum. A
mass of $1911\pm4$ MeV and a width of $29\pm11$ MeV are obtained.  

A state of similar mass has been also searched in  $\overline{p}d\rightarrow
p6\pi$ \cite{Gaspero}, and recently in $\overline{n}p$ annihilation in 
6 pions
by Obelix at LEAR \cite{Obelix}.
No clear signal is found in this mass region. We notice that a signal in
nucleon antinucleon annihilation could be interpreted as an indication for a
baryonium state. 


\subsection{Threshold enhancement in $J/\psi \rightarrow p\overline{p}\gamma$}
\begin{figure}[htb]
\centering
\includegraphics[width=7cm]{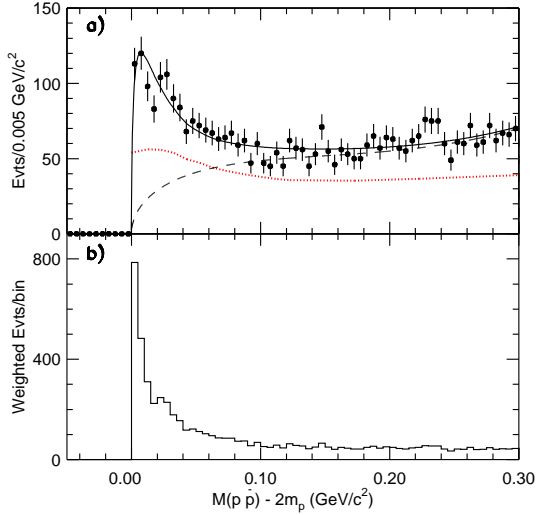}
 \caption{\it
(a) $M(p\overline{p})-2M_p$ distribution near threshold for $p\overline{p}\gamma$ 
events with fit superimposed. 
(b) Same distribution corrected for phase-space. Data from BES.  
    \label{BES1} }
\end{figure}

Very recently the BES collaboration at the $e^+e^-$ collider BEPC, Bejing, has
reported a study of the decay $J/\psi\rightarrow p\overline{p}\gamma$
\cite{Bes2}.
The invariant mass spectrum of the $p\overline{p}$ system is shown in
Fig.\ref{BES1}. It shows a clear enhancement very close to threshold.

$p\overline{p}\gamma$ final states could be due either to $J/\psi$ radiative
decay (in this case the $p\overline{p}$ system is expected to have scalar or
pseudo-scalar quantum numbers) or to radiative return on $e^+e^-\rightarrow
p\overline{p}$ due to the time-like proton form factors (vector quantum
numbers). BES angular analysis
of the events close to threshold suggests radiative decays rather than
radiative return.

BES has tried a fit in both scalar and pseudo-scalar state hypotheses. The values
obtained for the masses are: $1859^{+3}_{-10}$ MeV (pseudo-scalar hypothesis) 
and $1867.4\pm0.9$ MeV (scalar hypothesis). In both cases the state is very
narrow: $\Gamma=0\pm21$ MeV (pseudo-scalar hypothesis) and $4.6\pm 1.8$ MeV
(scalar hypothesis).

\subsection{Summary}
The states observed by E687 and by BES
are apparently of different quantum numbers and of
significantly different mass. So they have a different origin. 
Several experiments can search for these states: FOCUS can statistically 
improve the E687 signal; 
BABAR and BELLE can access to 6 pions and $p\overline{p}$ production
through initial state radiation, and study exclusive $B$ decays with
$p\overline{p}$ in the final state \cite{Belle}; finally 
CLEO-C will study $J/\psi$ decays.
We notice that a mass value of 1.9 GeV, other than close to the
nucleon-antinucleon threshold, is also the typical predicted value of the mass
of the hybrid states in the context of the tube-flux model \cite{Isgur1}.

\section{Gluonium content of $\eta'$}
The pseudo-scalar meson $\eta'(958)$ is considered a 
possible glueball candidate, or at least a $q\overline{q}$ state strongly
mixed with a glueball. The $\eta'$ wave function can be written
as: 
$$|\eta'>=X_{\eta'}|(u\overline{u}+d\overline{d})/\sqrt{2}>+Y_{\eta'}|s\overline{s}>
+ Z_{\eta'}|{\rm glueball}>$$
Since the $\phi$ is an almost pure $s\overline{s}$ state, the decay
$\phi\rightarrow\eta'\gamma$ selects the $Y_{\eta'}$ component of the $\eta'$
wave function. 

\begin{figure}[htb]
\centering
\includegraphics[width=8cm]{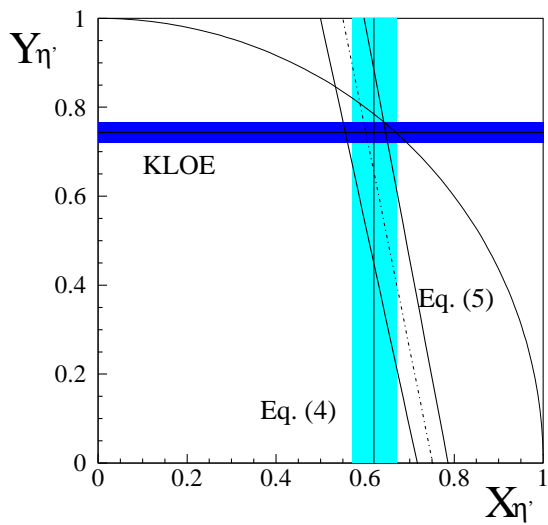}
 \caption{\it Bounds in the $X_{\eta'}$-$Y_{\eta'}$ plane from experimental
   branching fractions and SU(3) based calculations.
    \label{Gluetap} }
\end{figure}

KLOE has measured this decay with improved precision respect to previous
measurements \cite{KLOEetap}. 
If $Z_{\eta'}=0$, one has $X_{\eta'}^2+Y_{\eta'}^2=1$. The last
condition can be checked using together 
the $Y_{\eta'}$ value extracted by KLOE and two
other results on $\eta'$ decays\footnote{The two vertical bands are obtained
  using the PDG values for
  $\Gamma(\eta'\rightarrow\rho\gamma)/\Gamma(\omega\rightarrow\pi^0\gamma)$
  and for
  $\Gamma(\eta'\rightarrow\gamma\gamma)/\Gamma(\pi^0\rightarrow\gamma\gamma)$.}
and putting them in the  graph shown in
Fig.\ref{Gluetap}. The crossing of the three bands is compatible with the
condition   $X_{\eta'}^2+Y_{\eta'}^2=1$. This means that at this level of
experimental accuracies, there is 
no indication of a gluonium content of the $\eta'$. From the directly measured
ratio of $\phi\rightarrow\eta\gamma$ and $\phi\rightarrow\eta'\gamma$,
assuming $Z_{\eta'}=0$ and $Z_{\eta}=0$ KLOE has
determined the pseudo-scalar mixing angle in the flavour basis $\phi_p$ to be:
$$\phi_p=(41.8^{+1.9}_{-1.6})^{\circ}$$

\section{Conclusions}

\section{Acknowledgements}
I wish to thank the organisers of the conference, in particular W.Lohmann and
F.Fabbri. Among the people who helped me in putting together all what I
presented, I wish to thank in particular 
G.Adams, R.Baldini, G.Dunwoodie, A.Dzierba,
A.Filippi, F.Harris, S.Malvezzi, J.Napolitano, S.Serednyakov, J.Shan,
E.P.Solodov and A.Zallo.

\end{document}